\documentclass[conference]{IEEEtran}
\IEEEoverridecommandlockouts

\usepackage{url}
\usepackage{cite}
\usepackage{amsmath,amssymb,amsfonts}
\usepackage{siunitx}
\usepackage{graphicx}
\graphicspath{ {./figs_screenshots} }
\usepackage[caption=false]{subfig}

\def\BibTeX{{\rm B\kern-.05em{\sc i\kern-.025em b}\kern-.08em
T\kern-.1667em\lower.7ex\hbox{E}\kern-.125emX}}

\usepackage{tikz}
\usetikzlibrary{
	arrows.meta,
	shapes.geometric,
}
\makeatletter
\tikzset{
	fitting node/.style={
			inner sep=0pt,
			fill=none,
			draw=none,
			reset transform,
			fit={(\pgf@pathminx,\pgf@pathminy) (\pgf@pathmaxx,\pgf@pathmaxy)}
		},
	reset transform/.code={\pgftransformreset}
}
\makeatother

\usepackage{pgfplots}
\pgfplotsset{compat=newest}
\pgfplotsset{plot coordinates/math parser=false}
\pgfplotsset{
	legend image with text/.style={
			legend image code/.code={%
					\node[anchor=center] at (0.3cm,0cm) {#1};
				},
			line width=1pt
		},
}

\usepackage[acronyms,nonumberlist,nopostdot,nomain,nogroupskip,acronymlists={hidden}]{glossaries}
\newacronym{hpbw}{HPBW}{Half Power Beamwidth}
\newacronym{tle}{TLE}{Two-Line Element}
\newacronym{rmse}{RMSE}{Root Mean Square Error}
\newacronym{aps}{APS}{Adaptive Pattern Search}
\newacronym{iss}{ISS}{International Space Station}
\newacronym{leo}{LEO}{Low Earth Orbit}
\newacronym{roc}{RoC}{Rate of Change}
\newacronym{ntn}{NTN}{Non-Terrestrial Network}
\newacronym{thz}{THz}{Terahertz}
\newacronym{isl}{ISL}{Inter-Satellite Link}
\newacronym{sthz}{sub-THz}{sub-Terahertz}
\newacronym{ecdf}{ECDF}{Empirical Comulative Density Function}
\newacronym{pe}{PE}{Pointing Error}
\newacronym{sgp4}{SGP4}{Simplified General Perturbations}

\glsdisablehyper

\newlength\fheight
\newlength\fwidth
\usepackage{booktabs}

\usepackage[capitalize]{cleveref}
\creflabelformat{equation}{#2#1#3}
\creflabelformat{figure}{#2#1#3}
\creflabelformat{table}{#2#1#3}
\creflabelformat{section}{#2#1#3}

\usepackage{textcomp}
\usepackage{gensymb}

\usepackage{xspace}
\newcommand{\target}{satellite\xspace}

\title{Pointing-Error-Induced Fading in an Open-Loop THz Uplink with Hardware Impairments}

\author{
\IEEEauthorblockN{
	    Pietro Brach del Prever,~%
        Paolo Testolina,~%
        Ahmad Masihi,~%
        Sergey Petrushkevich,~%
        \\
        Michele Polese,~%
        Tommaso Melodia,~%
        Josep M. Jornet~%
	}
    \thanks{
    }
}

\begin{document}

\maketitle

\begin{abstract}
	We analyze the open-loop mechanical tracking performance of a \gls{sthz} and \gls{thz} uplink communication system.
	These high-frequency bands enable multi-gigabit links through large bandwidths and narrow beams, but require precise pointing to overcome spreading loss.
	A tracking system can be used to orient horn antennas toward mobile targets. We develop a mathematical model that captures the mechanical dynamics of a real tracking system, which includes motion latency and acceleration and velocity limits, to quantify pointing errors during satellite passes and integrate these effects into the link budget.
	We evaluate the trade-offs between beam directionality and pointing tolerance across different \gls{leo} satellite trajectories and control strategies.
	The results link the hardware limitations to the communications performance, providing design guidelines for high-frequency \gls{ntn} uplink under practical mechanical constraints.
\end{abstract}

\begin{IEEEkeywords}
	Terahertz, Non-Terrestrial Network, Pointing error, Fading
\end{IEEEkeywords}

\begin{picture}(0,0)(-10,-370)
	\put(0,0){
		\put(0,0){\footnotesize
			This paper has been accepted at IEEE Military Communications Conference (IEEE MILCOM 2025).
		}
		\put(0,-10){
			\scriptsize \textcopyright~2025 IEEE. Personal use of this material is permitted. Permission from IEEE must be obtained for all other uses, in any current or future media, including}
		\put(0, -17){
			\scriptsize reprinting/republishing this material for advertising or promotional purposes, creating new collective works, for resale or redistribution to servers or lists,}
		\put(0, -24){
			\scriptsize or reuse of any copyrighted component of this work in other works.}
	}
\end{picture}

\glsresetall

\section{Introduction}

The rapid growth in wireless data demand, combined with the increasing deployment of \gls{leo} constellations, is driving interest in frequency bands that can support multi-gigabit links over large distances.
\Gls{sthz} and \gls{thz} bands offer large contiguous bandwidths, thus being promising candidates for future \gls{ntn}, including ground-to-satellite links for broadband and scientific applications~\cite{araniti2021toward,aliaga2024non}.

High-gain beams are essential to overcome the severe spreading loss in long-range links at those frequencies, but they also make system performance sensitive to misalignment. In the uplink (from the ground to \gls{leo} satellites), this sensitivity is amplified by two factors: the rapid apparent motion of the satellite relative to the ground terminal, and the lack of instantaneous feedback in some operational modes, such as open-loop pointing, where the satellite does not provide any beacon for the ground transmitter lock-in. Furthermore, atmospheric absorption and turbulence introduce additional losses and potential beam distortion that do not occur in inter-satellite scenarios.

This work addresses these challenges by focusing on open-loop tracking for \gls{sthz}/\gls{thz} uplink. A tracking system can be used to orient horn antennas toward mobile targets. We develop a mathematical model that captures the mechanical and control-system dynamics of a real tracking system, quantifies the resulting pointing error over realistic satellite passes, and incorporates these effects into the link budget. This approach allows us to connect hardware-level limitations directly to link-level performance metrics.

The main contributions of this paper are:
\begin{itemize}
	\item A motion model of the antenna mounting system, including moving latency, acceleration limits, and sampling interval effects, enabling the simulation of pointing error profiles for different satellite trajectories and their impact on the link budget.
	\item An evaluation of the trade-offs between beam directionality and pointing tolerance, supported by performance results for multiple pass geometries and mount control strategies.
\end{itemize}
The results provide insight into the design of robust high-frequency \gls{ntn} uplink where precise beam alignment must be achieved under practical mechanical and atmospheric constraints.

\section{Related Work}
\label{sec:related_work}

High-frequency links in the \gls{sthz} and \gls{thz} ranges continue to attract significant attention for \gls{ntn} applications due to the wide available bandwidth and the need for narrow, high-gain beams to mitigate severe spreading losses.
In such systems, even minor \glspl{pe}, like satellite vibrations~\cite{ding2017analysis}, can drastically reduce received power.
Prior analyses of \gls{thz} inter-satellite links---combining geometric spreading and pointing loss models---highlight the interaction between beam width and pointing accuracy, and evaluate outage probabilities across frequencies and link distances~\cite{masihiISL}. Although focusing on near-vacuum inter-satellite environments, those methodologies and insights on beam tolerance remain highly applicable to terrestrial uplink scenarios.

The speed and accuracy of the ground terminal steering are equally critical. Foundational studies on ground-station tracking document challenges such as the azimuth/elevation ``zenith blind spot'' and the resulting demand for rapid azimuthal drive rates, directly impacting mount design, servo dynamics, and sampling fidelity~\cite{lozier1963servo,carlson1965coverage,borkowski1987nearzenith}.
In open-loop configurations, pointing error is fundamentally constrained by orbit-prediction reliability (e.g., \gls{sgp4} based on publicly available \glspl{tle} files), with real-world assessments reporting error evolution over prediction horizons~\cite{vallado2006revisiting,aida2013sgp4}.

Adding to this foundation, classical surveys like Hawkins et al. (1988) comprehensively review antenna tracking methodologies for satellite communications, discussing orbit determination, estimation approaches, control strategies, and practical deployment considerations, including intelligent control algorithms for tracking terrestrial and spaceborne links~\cite{hawkins1988tracking}. Expanding focus beyond fixed ground stations, Kim et al. (2013) propose a two-axis gyrosensor-based BLDC motor control system configured for parabolic antennas on mobile platforms—capable of compensating six-degree-of-freedom motion (yaw, pitch, roll) using encoder feedback and AGC-guided tracking~\cite{kim2013precise}. Complementing these, Larsen et al. (2025) explore an optical approach: ground-based telescopes observe a CubeSat LED payload to facilitate optical tracking in congested RF environments, showcasing an alternative strategy for maintaining, or verifying, satellite pointing in scenarios where RF links are constrained~\cite{larsen2025optical}.

In this work, we adapt the \gls{thz} analytical framework of~\cite{masihiISL} for uplink configurations, embedding realistic ground-terminal pointing constraints, and evaluating performance across typical \gls{leo} link ranges.

\section{System Description}\label{sec:system}

In this work, we consider a ground-to-satellite \gls{thz} uplink.
The ground transmitter antenna is hosted on a motorized system controlled by an electronic unit.
The system works in open loop, i.e., the satellite does not provide a beacon for the ground transmitter to lock in.
In this case, the beam alignment procedure can only rely on the trajectory of the satellite, either estimated by feeding the \glspl{tle} data to an orbit propagator or provided by the satellite via a secondary link.
For simplicity, in this work we assume that the ground controller has perfect knowledge of the satellite trajectory, i.e., we neglect the pointing error due to approximations of the orbital propagator and errors in the \gls{tle}.

\subsection{Hardware System Description}\label{sec:hardware_system_description}

We consider an \textit{alt-azimuth mount}.
An alt-azimuth mount is a mechanical mounting system commonly used for antennas or telescopes that can be rotated around two perpendicular axes, a horizontal one (altitude), which varies the elevation, and a vertical one, which varies the azimuth. Azimuth and elevation are the spherical coordinates in the mount reference system, centered at the intersection of the rotation axes.
In the rest of the paper we refer to a generic angular coordinate $\alpha$, as the following formulation is equivalent for azimuth and elevation.
Specifically, $\alpha_{sat}$ indicates the \target coordinates and $\alpha_{M}$ indicates the mount coordinates, i.e., where the antenna is pointing.

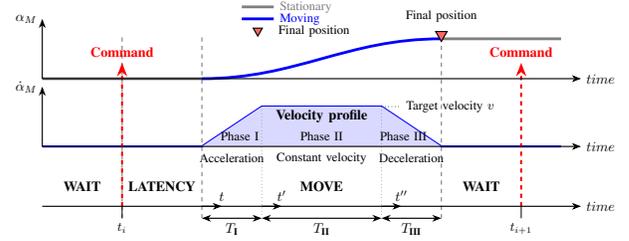
\begin{figure}[t]
	\centering
	\setlength{\fwidth}{0.95\columnwidth}
	\resizebox{0.98\fwidth}{!}{
\begin{tikzpicture}[
   >=Stealth,
   final/.style={star, star points=3, star point ratio=0.5, draw, thick, fill=red!60},
   command/.style={->, thick, red, dashed},
   timeline/.style={thick, black},
   annotation/.style={font=\small}
]

\draw[blue, line width=1.5pt] (0,1.5) -- (4,1.5) -- (5.5,2.5) -- (8.5,2.5) -- (10,1.5) -- (13,1.5);
\fill[blue!15] (4,1.5) -- (5.5,2.5) -- (8.5,2.5) -- (10,1.5) -- cycle;

\node[annotation] at (4.95,1.75) {Phase I};
\node[annotation] at (7,1.75) {Phase II};
\node[annotation] at (9.05,1.75) {Phase III};
\node[annotation] at (4.75,1.2) {Acceleration};
\node[annotation] at (7,1.2) {Constant velocity};
\node[annotation] at (9.25,1.2) {Deceleration};
\draw[dotted] (8.5,2.5) -- (9,2.5) node[annotation, right] {Target velocity $v$};

\draw[gray, dashed, thick] (2,-0.3) -- (2,4.3);
\draw[gray, dashed, thick] (4,-0.3) -- (4,4.3);
\draw[gray, dashed, thick] (10,-0.3) -- (10,4.3);
\draw[gray, dotted] (5.5,-0.3) -- (5.5,2.5);
\draw[gray, dotted] (8.5,-0.3) -- (8.5,2.5);

\node[annotation, font=\bfseries] at (7,2.25) {Velocity profile};

\draw[line width=2pt, gray] (0,3.2) -- (4,3.2);
\draw[line width=2pt, blue] (4,3.2) to[out=0, in=180] (10,4.2);
\draw[line width=2pt, gray] (10,4.2) -- (13,4.2);

\node[final] at (10,4.3) {};
\node[annotation, above] at (10,4.5) {Final position};

\begin{scope}[shift={(5,5)}]
   \draw[line width=2pt, gray] (0,0) -- (0.8,0) node[right, annotation] {Stationary};
   \draw[line width=2pt, blue] (0,-0.3) -- (0.8,-0.3) node[right, annotation] {Moving};
   \node[final, scale=0.8] at (0.4,-0.6) {} node[right=0.4cm, annotation] at (0.4,-0.6) {Final position};
\end{scope}

\draw (2,-.3) -- (2,0.0) node[annotation, below=3mm] {$t_{i}$};
\draw (12,-.3) -- (12,0.0) node[annotation, below=3mm] {$t_{i+1}$};

\draw[command, line width=1.5pt] (2,0) -- (2,3.6) node[above, annotation, font=\bfseries] {Command};
\draw[command, line width=1.5pt] (12,0) -- (12,3.6) node[above, annotation, font=\bfseries] {Command};

\draw[timeline, ->] (0,1.5) -- (13.5,1.5) node[right] {$time$};
\draw[timeline, ->] (0,1.5) -- (0,3) node[left] {$\dot{\alpha}_M$};

\draw[timeline, ->] (0,3.2) -- (13.5,3.2) node[right] {$time$};
\draw[timeline, ->] (0,3.2) -- (0,4.7) node[left] {$\alpha_M$};

\draw[timeline, ->] (0,0) -- (13.5,0) node[right] {\textbf{$time$}};
\draw[timeline, ->] (4,0.0) -- (4.5,0.0) node[above] {\textbf{$t$}};
\draw[timeline, ->] (5.5,0.0) -- (6,0.0) node[above] {\textbf{$t'$}};
\draw[timeline, ->] (8.5,0.0) -- (9,0.0) node[above] {\textbf{$t''$}};
\draw[timeline, <->] (4,-0.3) -- node[below] {\textbf{$T_{\text{I}}$}} (5.5,-0.3);
\draw[timeline, <->] (5.5,-0.3) -- node[below] {\textbf{$T_{\text{II}}$}} (8.5,-0.3);
\draw[timeline, <->] (8.5,-0.3) -- node[below] {\textbf{$T_{\text{III}}$}} (10,-0.3);

\node[font=\bfseries] at (1,0.5) {WAIT};
\node[font=\bfseries] at (3,0.5) {LATENCY};
\node[font=\bfseries] at (7,0.5) {MOVE};
\node[font=\bfseries] at (11,0.5) {WAIT};

\end{tikzpicture}}
	\caption{Phases of movement at each time step. Velocity and position refer to a generic angular coordinate.}
	\label{fig:mount_scheme}
\end{figure}

We make the following assumptions.
\begin{enumerate}
	\item The rotation axes cannot be controlled independently: this is the case, for instance, when the motors are connected in series.\label{ass:not_independent}
	\item Commands sent to the mount are executed after a non-negligible latency $l$ (LATENCY phase in~\cref{fig:mount_scheme}).\label{ass:latency}
	\item The mount does not have a buffer, thus, commands cannot be queued and scheduled at the mount, i.e., each command introduces latency $l$ and that cannot be reduced.\label{ass:no_buffer}
	\item We assume that commands can be sent to the mount at uniform time intervals $\Delta t_i = t_{i+1}-t_i=\Delta t$, $i\in \mathbb{N}$. This is a simplified yet realistic assumption due to Assumption~\ref{ass:no_buffer}, which requires the commands to be continuously issued, introducing the corresponding latency (Assumption~\ref{ass:latency}), and due to the dual axes control (Assumption~\ref{ass:not_independent}), which complicates the selection of an optimal time step due to different characteristics of the \target trajectory in the azimuth and elevation axis.
	\item The velocity cannot be dynamically adjusted due to latency constraints or hardware limitations. Thus, commands only transmit to the mount the position where to move.
	\item We assume $\Delta t$ to be large enough for the command to be received by the mount and executed.
\end{enumerate}

Thus, after a controller command is issued, the mount enters the LATENCY phase, where the command indicating the position to reach for the current step is sent to the mount, as shown in~\cref{fig:mount_scheme}. Finally, the mount moves, subsequently for each axis, to the desired position during the mount movement (MOVE phase).

As shown in~\cref{fig:mount_scheme}, we assume the law of motion during the MOVE phase to follow a trapezoidal velocity profile, that defines the motion across three successive phases: (I) one of constant acceleration, (II) one of constant velocity, and (III) one of constant deceleration.
The acceleration $\ddot{\alpha}_M(t)$ is a piecewise function, defined, for each axis, as
\begin{equation} \label{eq:mount_accel}
	\ddot{\alpha}_M(t) =
	\begin{cases}
		+a & t\in\text{Phase (I): Acceleration}       \\
		0  & t\in\text{Phase (II): Constant velocity} \\
		-a & t\in\text{Phase (III): Deceleration}     \\
	\end{cases},
\end{equation}
where $a$ is the selected acceleration, and $t \in [0,T]$.

While this model involves theoretical discontinuities in acceleration (infinite jerk) at the transitions between phases, the practical impact on a well-damped mechanical system is often minimal and contained within the overall error budget.
We assume that the structural integrity and control algorithms of the mount are designed to handle these transitions without inducing significant detrimental vibration.

$\alpha_{M,0}$ is the mount positions at the beginning of the movement,
$v$ is the constant target velocity to reach during Phase (II),
$T_{\text{I}} = v / a$ is the time the mount accelerates in Phase (I) or decelerates in Phase (III),
$T_{\text{II}} = T - 2 \cdot T_{\text{I}}$ is the time the mount moves at constant speed in Phase (II),
and $t' = t - T_{\text{I}}$ and
$t'' = t - T_{\text{I}} - T_{\text{II}}$ are is time elapsed from the beginning of Phase (II) and (III), respectively.
Then, integrating~\cref{eq:mount_accel}, the velocity $\dot{\alpha}_M(t)$ and position $\alpha_{M}(t)$ of the mount are
\begin{equation}\label{eq:mount_velocity}
	\dot{\alpha}_{M}(t) = \int_{0}^{T} \ddot{\alpha}_{M} \,dt =
	\begin{cases}
		a \cdot t       & t\in\text{(I)}   \\
		v               & t\in\text{(II)}  \\
		v - a \cdot t'' & t\in\text{(III)}
	\end{cases},
\end{equation}
\begin{multline}\label{eq:mount_position}
	\alpha_{M}(t) = \int_{0}^{T} \dot{\alpha}_{M} \,dt = \\
	=\begin{cases}
		\alpha_{M,0} + \frac{1}{2}a \cdot t^2                                                        & t\in\text{(I)}   \\
		\alpha_{M,0} + \frac{1}{2}a T_{\text{I}}^2 + v \cdot t'                                      & t\in\text{(II)}  \\
		\alpha_{M,0} + \frac{1}{2}a T_{\text{I}}^2 + vT_{\text{II}} + (v - \frac{1}{2}a) \cdot t''^2 & t\in\text{(III)}
	\end{cases}.
\end{multline}

\subsection{Antenna Model and Pointing Error Loss}
In this work, we consider an uplink scenario consisting of a transmitter located on the ground and a receiver mounted on a small \gls{leo} satellite.
Directional antennas with beam widths of a few degrees or less are assumed to be used at both the transmitter and the receiver.
The received power $P_r$ at the satellite can be calculated from
\begin{equation}\label{eq:Pr_no_error}
	P_r =  \frac{P_t}{L_{S}L_{a}L_{p}},
\end{equation}
where $P_t$ is the transmitted power, respectively, and $L_s$, $L_a$, and $L_p$ are the spreading loss, absorption loss, and pointing loss, respectively. In this study, we focus only on pointing loss, aiming to quantify how much misalignment can reduce the received power.
If the distance between the transmitter and receiver is $d$, the wavelength of the beam is $\lambda = c/f$, where $f$ is the carrier frequency, and $G_t$ and $G_r$ are the transmitter and receiver antenna gains, respectively.
When $8\sqrt{\pi}d \gg \sqrt{G_t.G_r}\lambda$, the expression for pointing loss in~\cite{masihiISL} can be simplified to:
\begin{equation}\label{eq:Lp}
	L_{p} =\exp \left( \frac{\pi \eta_a A_t}{\lambda^2}\tan^2\alpha_E \right),
\end{equation}
where $\eta_a$ is the antenna aperture efficiency, $A_t$ is the physical aperture area of the antenna, and $\alpha_E$ is the pointing error.
To give a sense of the antenna size at different \gls{sthz} frequencies,
consider an antenna with a gain of $51$~dBi and aperture efficiency $\eta = 70\%$.
At D-band ($130$~GHz), the corresponding physical aperture is approximately
$241$~cm$^2$. At $220$~GHz, the required aperture
shrinks to $84$~cm$^2$, while at $660$~GHz the same gain
requires less than $9.4$~cm$^2$ antenna aperture area.
Similarly, for a $60$~dBi antenna, the equivalent apertures are $6054$~cm$^2$, $2113$~cm$^2$, and $235$~cm$^2$, respectively.

This analysis is carried out under the assumption that the receiver antenna on the satellite is perfectly oriented toward the ground station.

\section{Problem Formulation}

\subsection{Pointing Error}
\label{sec:pointing_error}
The pointing error $\alpha_{E}$ between the $\alpha_{sat}$ and $\alpha_{M}$ is computed using and the Haversine angular distance between two positions in spherical coordinates:
\begin{multline}\label{eq:haversine}
	\alpha_{E} = \\
	\arcsin\sqrt{\sin^2\left(\frac{\Delta\theta}{2}\right) + \cos(\theta_M) \cos(\theta_{sat}) \sin^2\left(\frac{\Delta\phi}{2}\right)},
\end{multline}
where $\theta_M$ and $\theta_{sat}$ represent the elevation of the mount and of the \target, respectively, $\Delta\theta$ their difference, and $\Delta\phi$ the difference between their azimuth.
This formula calculates the angular distance between two points in space as a function of their latitude and longitude, and it is numerically stable for both small and large distances.

\subsection{Velocity Control}\label{sec:velocity_optimization}
For the mount control we consider a look-ahead approach, i.e., the commands are transmitted to the mount so that it reaches the next trajectory position before the \target.

To do so, we introduce a WAIT phase in each time step, as reported in \cref{fig:mount_scheme}: the controller waits for the \target to reach the mount position before issuing the next command to reach the next position.

For the evaluation of the pointing error loss, we consider three different velocity profiles.
For simplicity, we tune only the target velocity of the trapezoidal profile described in~\cref{sec:hardware_system_description}: varying the acceleration impacts the jerk of the mount, which introduces higher-order effects that are beyond the scope of this paper.

We consider two naive approaches for baseline: setting the target velocity of the trapezoidal profile equal to the  maximum velocity allowed by the mount (Profile A) and to the maximum velocity of the \target during the considered trajectory (Profile B) considered.

Finally, we consider the target velocity (Profile C) obtained using \gls{aps}, a derivative-free optimization algorithm that explores the solution space through coordinate searches with dynamically adjusted step sizes.
This approach is well-suited for this problem, where the objective function, minimizing the \gls{rmse}, is convex but non-differentiable.
\gls{aps} provides deterministic, reproducible optimization while effectively navigating the non-smooth objective landscape created by velocity limitations, acceleration constraints, and trapezoidal motion profiles.
The algorithm explores four coordinate directions (steps of $\pm\SI{2}{\degree\per\second}$ for azimuth and elevation velocities), accepting improvements and halving the step size when no progress is found.
Optimization terminates when the step size falls below $\SI{0.1}{\degree\per\second}$,
or after 20 iterations.

\section{Results}
For the evaluation of the simulator, of the velocity optimization, and for the analysis of the pointing loss, we consider the \gls{iss} as representative of a \gls{leo} satellite.
We selected three transits over Boston, MA, USA (Lat: $\ang{42.3601}$ latitude, Lon: $\ang{-71.0589}$), between August 10th and 12th, 2025, with different peak elevation angles: $\ang{47}$, $\ang{70}$, and $\ang{83}$.
The trajectory is computed based on the \gls{tle} data from the Celestrak catalog.\footnote{\url{https://celestrak.org/}}
In the remainder of this paper, the trajectories are identified and referred to with their peak elevation.

\subsection{Pointing Error Analysis}
\label{sec:pe_analysis}

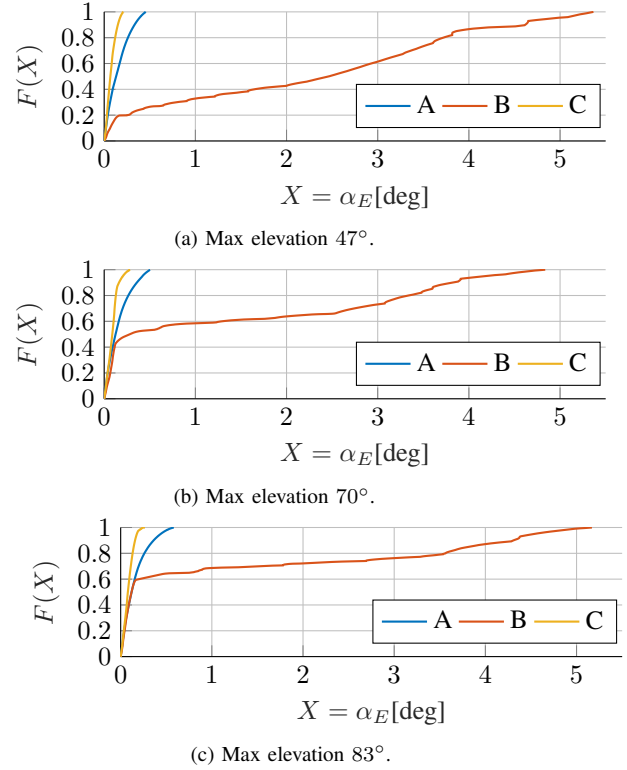
\begin{figure}[t]
	\centering
	\setlength{\fheight}{.2\columnwidth}
	\setlength{\fwidth}{0.8\columnwidth}

	\hspace{-0.1\columnwidth}\subfloat[Max elevation $\ang{47}$.]{
		\centering
		\definecolor{matlabblue}{rgb}{0.0000, 0.4470, 0.7410}
\definecolor{matlaborange}{rgb}{0.8500, 0.3250, 0.0980}
\definecolor{matlabyellow}{rgb}{0.9290, 0.6940, 0.1250}
\definecolor{matlabpurple}{rgb}{0.4940, 0.1840, 0.5560}
\definecolor{matlabgreen}{rgb}{0.4660, 0.6740, 0.1880}
\definecolor{matlablightblue}{rgb}{0.3010, 0.7450, 0.9330}
\definecolor{matlabred}{rgb}{0.6350, 0.0780, 0.1840}

\begin{tikzpicture}

\begin{axis}[%
width=0.951\fwidth,
height=0.98\fheight,
at={(0\fwidth,0\fheight)},
scale only axis,
xlabel style={font=\color{white!15!black}},
xlabel={$X = \alpha_E$[deg]},
xmin=0,
xmax=5.5,
ymin=0,
ymax=1,
ylabel style={font=\color{white!15!black}},
ylabel={$F(X)$},
axis background/.style={fill=white},
axis x line*=bottom,
axis y line*=left,
xmajorgrids,
ymajorgrids,
legend style={
    legend columns=3,
    at={(0.99,0.1)},
    anchor=south east,
},
]
\addlegendentry{A}
\addlegendentry{B}
\addlegendentry{C}

\addplot[no markers,matlabblue, solid, line width = 0.8] table [x=x, y=F, col sep=comma] {figs/data/pe_ecdf/elmax47_case_A_ecdf.csv};

\addplot[no markers,matlaborange, solid, line width = 0.8] table [x=x, y=F, col sep=comma] {figs/data/pe_ecdf/elmax47_case_B_ecdf.csv};

\addplot[no markers,matlabyellow, solid, line width = 0.8] table [x=x, y=F, col sep=comma] {figs/data/pe_ecdf/elmax47_case_C_ecdf.csv};

\end{axis}
\end{tikzpicture}%
	}

	\hspace{-0.1\columnwidth}\subfloat[Max elevation $\ang{70}$.]{
		\centering
		\definecolor{matlabblue}{rgb}{0.0000, 0.4470, 0.7410}
\definecolor{matlaborange}{rgb}{0.8500, 0.3250, 0.0980}
\definecolor{matlabyellow}{rgb}{0.9290, 0.6940, 0.1250}
\definecolor{matlabpurple}{rgb}{0.4940, 0.1840, 0.5560}
\definecolor{matlabgreen}{rgb}{0.4660, 0.6740, 0.1880}
\definecolor{matlablightblue}{rgb}{0.3010, 0.7450, 0.9330}
\definecolor{matlabred}{rgb}{0.6350, 0.0780, 0.1840}

\begin{tikzpicture}

\begin{axis}[%
width=0.951\fwidth,
height=0.98\fheight,
at={(0\fwidth,0\fheight)},
scale only axis,
xlabel style={font=\color{white!15!black}},
xlabel={$X = \alpha_E$[deg]},
xmin=0,
xmax=5.5,
ymin=0,
ymax=1,
ylabel style={font=\color{white!15!black}},
ylabel={$F(X)$},
axis background/.style={fill=white},
axis x line*=bottom,
axis y line*=left,
xmajorgrids,
ymajorgrids,
legend style={
    legend columns=3,
    at={(0.99,0.1)},
    anchor=south east,
},
]
\addlegendentry{A}
\addlegendentry{B}
\addlegendentry{C}

\addplot[no markers,matlabblue, line width = 0.8] table [x=x, y=F, col sep=comma] {figs/data/pe_ecdf/elmax70_case_A_ecdf.csv};

\addplot[no markers,matlaborange, line width = 0.8] table [x=x, y=F, col sep=comma] {figs/data/pe_ecdf/elmax70_case_B_ecdf.csv};

\addplot[no markers,matlabyellow, line width = 0.8] table [x=x, y=F, col sep=comma] {figs/data/pe_ecdf/elmax70_case_C_ecdf.csv};

\end{axis}
\end{tikzpicture}%
	}

	\hspace{-0.05\columnwidth}\subfloat[Max elevation $\ang{83}$.]{
		\centering
		\definecolor{matlabblue}{rgb}{0.0000, 0.4470, 0.7410}
\definecolor{matlaborange}{rgb}{0.8500, 0.3250, 0.0980}
\definecolor{matlabyellow}{rgb}{0.9290, 0.6940, 0.1250}
\definecolor{matlabpurple}{rgb}{0.4940, 0.1840, 0.5560}
\definecolor{matlabgreen}{rgb}{0.4660, 0.6740, 0.1880}
\definecolor{matlablightblue}{rgb}{0.3010, 0.7450, 0.9330}
\definecolor{matlabred}{rgb}{0.6350, 0.0780, 0.1840}

\begin{tikzpicture}

\begin{axis}[%
width=0.951\fwidth,
height=0.98\fheight,
at={(0\fwidth,0\fheight)},
scale only axis,
xmin=0,
xlabel style={font=\color{white!15!black}},
xlabel={$X = \alpha_E$[deg]},
xmin=0,
xmax=5.5,
ymin=0,
ymax=1,
ylabel style={font=\color{white!15!black}},
ylabel={$F(X)$},
axis background/.style={fill=white},
axis x line*=bottom,
axis y line*=left,
xmajorgrids,
ymajorgrids,
legend style={
    legend columns=3,
    at={(0.99,0.1)},
    anchor=south east,
},
]
\addlegendentry{A}
\addlegendentry{B}
\addlegendentry{C}

\addplot[no markers,matlabblue, line width = 0.8] table [x=x, y=F, col sep=comma] {figs/data/pe_ecdf/elmax83_case_A_ecdf.csv};

\addplot[no markers,matlaborange, line width = 0.8] table [x=x, y=F, col sep=comma] {figs/data/pe_ecdf/elmax83_case_B_ecdf.csv};

\addplot[no markers,matlabyellow, line width = 0.8] table [x=x, y=F, col sep=comma] {figs/data/pe_ecdf/elmax83_case_C_ecdf.csv};

\end{axis}
\end{tikzpicture}%
	}
	\caption{\gls{ecdf} of the pointing error for the three considered trajectories with maximum altitude $\ang{47}$, $\ang{70}$, and $\ang{83}$, and  the three considered velocity profiles A, B, and C.}
	\label{fig:pe_ecdf}
\end{figure}

In the simulation, we used typical hardware parameters (response delay, maximum velocity, and acceleration, reported in~\cref{tab:simulation_parameters})
for a mid-range alt-azimuth mechanical mount.
The simulation step is $\SI{5}{\milli\second}$.
\Cref{tab:simulation_velocities} shows the velocity set for each trajectory and velocity profile.
\Cref{fig:pe_ecdf} reports results of the simulations: for each trajectory and strategy the \gls{ecdf} of the pointing error is shown.

\begin{table}[t]
	\centering
	\caption{System parameters used in the simulation. The velocity profile settings are the same for azimuth and elevation.}
	\label{tab:simulation_parameters}
	\begin{tabular}{cc}\toprule
		Parameter                        & Value                         \\\midrule
		Velocity Profile                 & Trapezoidal                   \\
		Maximum velocity                 & \SI{10}{\degree\per\second}   \\
		Acceleration                     & \SI{20}{\degree\per\second^2} \\
		Command interval time $\Delta t$ & \SI{1}{\second}               \\
		Response delay (LATENCY)         & \SI{100}{\milli\second}       \\\bottomrule
	\end{tabular}
\end{table}

\begin{table}[t]
	\centering
	\caption{Target velocity (azimuth/elevation) for each velocity profile and \target trajectory.}
	\label{tab:simulation_velocities}
	\begin{tabular}{cccc}
		\toprule
		           & A (az./el.) [$\si{\degree}/\si{\second}$] & B (az./el.) [$\si{\degree}/\si{\second}$] & C (az./el.) [$\si{\degree}/\si{\second}$] \\
		\midrule
		$\ang{48}$ & 10.0/10.0                                 & 1.1/0.3                                   & 1.3/0.3                                   \\
		$\ang{70}$ & 10.0/10.0                                 & 2.8/0.6                                   & 3.3/0.7                                   \\
		$\ang{84}$ & 10.0/10.0                                 & 9.1/0.9                                   & 10.0/2.0                                  \\
		\bottomrule
	\end{tabular}
\end{table}

\begin{figure}[t]
	\centering
	\setlength{\fheight}{0.28\columnwidth}
	\definecolor{matlabblue}{rgb}{0.0000, 0.4470, 0.7410}
\definecolor{matlaborange}{rgb}{0.8500, 0.3250, 0.0980}
\definecolor{matlabyellow}{rgb}{0.9290, 0.6940, 0.1250}
\definecolor{matlabpurple}{rgb}{0.4940, 0.1840, 0.5560}
\definecolor{matlabgreen}{rgb}{0.4660, 0.6740, 0.1880}
\definecolor{matlablightblue}{rgb}{0.3010, 0.7450, 0.9330}
\definecolor{matlabred}{rgb}{0.6350, 0.0780, 0.1840}

\begin{tikzpicture}

\begin{axis}[%
width=0.951\fwidth,
height=\fheight,
at={(0\fwidth,0\fheight)},
scale only axis,
xlabel style={font=\color{white!15!black}},
xlabel={Time [\si{\second}]},
ylabel style={font=\color{white!15!black}},
ylabel={Azimuth [deg]},
axis background/.style={fill=white},
axis x line*=bottom,
axis y line*=left,
xmin=0,
xmax=240,
xmajorgrids,
ymajorgrids,
legend style={
    at={(0.02,0.02)},
    anchor=south west,
    font=\footnotesize
},
enlargelimits=false,
]

\addplot[
    color=matlabblue,
    no markers,
    line width=1pt,
] table [x=time, y=target, col sep=comma] {figs/data/lag/elmax83_iss_case_B_azimuth_downsampled.csv};
\addlegendentry{Satellite}

\addplot[
    color=matlablightblue,
    no markers,
    line width=1pt,
    dashdotted
] table [x=time, y=mount, col sep=comma] {figs/data/lag/elmax83_iss_case_B_azimuth_downsampled.csv};
\addlegendentry{Vel. Profile B}

\end{axis}

\end{tikzpicture}%
	\caption{Lagging behavior of velocity profile B for the $\ang{83}$ trajectory.}
	\label{fig:lag}
\end{figure}
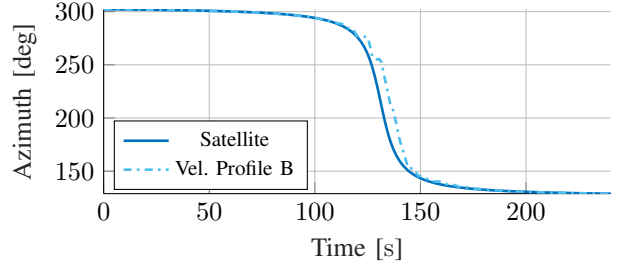

Strategy B has the worst performance among the three velocity profiles, with only $40\%$, $60\%$, and $70\%$ of samples within $\ang{1}$ of pointing error for the three trajectories, respectively.
This is because at each timestep, the mount movement is delayed in the LATENCY phase, and its average speed is not enough to keep up with the satellite velocity exactly where the satellite itself reaches its maximum velocity, thus invalidating our model assumption that the mount always reaches the target position before the satellite leaves it. This behavior is evident from~\cref{fig:lag}.

Strategy A performs much better, with most data points having an error below the $\ang{0.5}$ for all three strategies. This is a good result and heavily depends on the maximum values of velocity set for each axis (azimuth and elevation).
Thus, we deduce that for the system to keep up with the satellite movement, the mount velocity needs to be higher than the maximum satellite velocity, especially in the presence of significant hardware latency.

Strategy C represents the optimized velocity value for each axis.
It shows improved performance compared to strategy A, while keeping the values of velocity for each axis to a value lower than those for velocity profile A.

\begin{figure}[t]
	\centering
	\setlength{\fheight}{0.3\columnwidth}
	\definecolor{matlabblue}{rgb}{0.0000, 0.4470, 0.7410}
\definecolor{matlaborange}{rgb}{0.8500, 0.3250, 0.0980}
\definecolor{matlabyellow}{rgb}{0.9290, 0.6940, 0.1250}
\definecolor{matlabpurple}{rgb}{0.4940, 0.1840, 0.5560}
\definecolor{matlabgreen}{rgb}{0.4660, 0.6740, 0.1880}
\definecolor{matlablightblue}{rgb}{0.3010, 0.7450, 0.9330}
\definecolor{matlabred}{rgb}{0.6350, 0.0780, 0.1840}

\begin{tikzpicture}
\begin{axis}[%
width=0.8\fwidth,
height=\fheight,
at={(0\fwidth,0\fheight)},
scale only axis,
xlabel style={font=\color{white!15!black}},
xlabel={Time [\si{\second}]},
ylabel style={font=\color{white!15!black}},
ylabel={Azimuth [deg]},
ylabel style={color=matlabblue},
axis background/.style={fill=white},
axis x line*=bottom,
axis y line*=left,
xmajorgrids,
ymajorgrids,
legend columns = 3,
legend style={
    at={(0.5,1.05)},
    anchor=south,
    font=\footnotesize,
},
enlargelimits=false,
xmin=122.7,
xmax=127.8,
]
\addlegendimage{color=black,no markers,line width=.8pt}
\addlegendentry{Satellite}
\addlegendimage{color=black,no markers,line width=1pt,dashed}
\addlegendentry{Vel. Profile A}
\addlegendimage{color=black,no markers,line width=1.5pt,dotted}
\addlegendentry{Vel. Profile C}

\addplot[
    color=matlabblue,
    no markers,
    line width=.8pt,
] table [x=time_seconds, y=az_target, col sep=comma] {figs/data/oscillation_A_47.csv};

\addplot[
    color=matlabblue,
    line width=1pt,
    dashed,
] table [x=time_seconds, y=az_mount, col sep=comma] {figs/data/oscillation_A_47.csv};

\addplot[
    color=matlabblue,
    line width=1.5pt,
    dotted
] table [x=x, y=y, col sep=comma] {figs/data/elmax47_case_C_t_vs_az_mount.csv};
\end{axis}

\begin{axis}[%
width=0.8\fwidth,
height=0.98\fheight,
at={(0\fwidth,0\fheight)},
scale only axis,
axis x line*=none, %
axis y line*=right,
y axis line style={matlabred},
y tick label style={matlabred},
ylabel={Error [deg]},
ylabel near ticks,
ylabel style={color=matlabred},
yticklabel pos=right,
ytick style={color=matlabred},
xtick=\empty, %
xticklabels={}, %
legend style={
    at={(0.98,1.0)},
    anchor=north east,
    font=\footnotesize
},
enlargelimits=false,
xmin=122.7,
xmax=127.8,
ymax=0.51,
ymin=0
]
\addplot[
    color=matlabred,
    dashed,
    line width=0.8pt,
] table [x=time_seconds, y=total_error, col sep=comma] {figs/data/oscillation_A_47.csv};
\addplot[
    color=matlabred,
    dotted,
    line width=1.5pt,
] table [x=x, y=y, col sep=comma] {figs/data/elmax47_case_C_t_vs_pe.csv};
\end{axis}
\end{tikzpicture}
	\caption{Oscillatory behavior of the mount with velocity profile A and C for $\ang{47}$ trajectory: the WAIT, LATENCY, and MOVE phases are clearly distinguishable for each movement.}
	\label{fig:oscillations}
\end{figure}
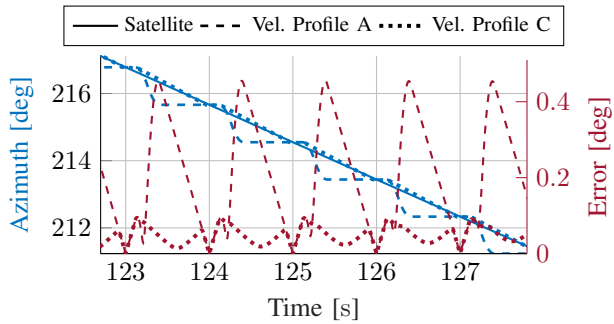

As shown in~\cref{fig:oscillations}, all strategies, including the optimized one, exhibit oscillatory behavior for the pointing error throughout the simulation.
The mount follows a periodic pattern: after the command latency (error increases during the LATENCY phase), it catches up with the \target (error decreases to zero), and reaches the next position (error increases), where it waits for the next command (error decreases).
As expected, setting the mount velocity to its maximum (profile A) exacerbates the oscillations, when it doesn't match the \target speed: the antenna reaches its next position much earlier than the \target.
Conversely, a mount moving at the same speed (profile B) as the \target always lags behind, due to the initial latency and the trapezoidal velocity profile.
When it is not possible to set the velocity of the mount at each time step, Profile C strikes a balance between the two strategies, significantly reducing the peak error.

\subsection{Pointing Error-Induced Fading}\label{sec:error_fading}

\begin{figure}[t]
	\centering
	\setlength{\fheight}{.24\columnwidth}
	\setlength{\fwidth}{0.8\columnwidth}

	\subfloat{
		\centering
		\setlength{\fheight}{.24\columnwidth}
		\setlength{\fwidth}{\columnwidth}
		\definecolor{matlabblue}{rgb}{0.0000, 0.4470, 0.7410}
\definecolor{matlaborange}{rgb}{0.8500, 0.3250, 0.0980}
\definecolor{matlabyellow}{rgb}{0.9290, 0.6940, 0.1250}
\definecolor{matlabpurple}{rgb}{0.4940, 0.1840, 0.5560}
\definecolor{matlabgreen}{rgb}{0.4660, 0.6740, 0.1880}
\definecolor{matlablightblue}{rgb}{0.3010, 0.7450, 0.9330}
\definecolor{matlabred}{rgb}{0.6350, 0.0780, 0.1840}

\begin{tikzpicture}

\begin{axis}[%
width=0.9125\fwidth,
height=0.9125\fheight,
at={(0\fwidth,0\fheight)},
scale only axis,
hide axis,
legend style={
    legend columns=3,
    /tikz/every even column/.append style={column sep=0mm}
        },
    xmin=10,
    xmax=50,
    ymin=0,
    ymax=0.4,
]

\addlegendimage{black, solid, line width = 0.8}
\addlegendentry{60 dBi/0.2$\degree$}
\addlegendimage{black, dashed, line width = 0.8}
\addlegendentry{51 dBi/0.6$\degree$}
\addlegendimage{black, dotted, line width = 0.8}
\addlegendentry{46 dBi/1$\degree$}
\addlegendimage{matlabblue, solid, line width = 0.8,  only marks}
\addlegendentry{A}
\addlegendimage{matlaborange, solid, line width = 0.8, only marks}
\addlegendentry{B}
\addlegendimage{matlabyellow, solid, line width = 0.8, only marks}
\addlegendentry{C}
\end{axis}

\end{tikzpicture}%
	}

	\hspace{-0.1\columnwidth}\subfloat[Max elevation $\ang{47}$.]{
		\centering
		\definecolor{matlabblue}{rgb}{0.0000, 0.4470, 0.7410}
\definecolor{matlaborange}{rgb}{0.8500, 0.3250, 0.0980}
\definecolor{matlabyellow}{rgb}{0.9290, 0.6940, 0.1250}
\definecolor{matlabpurple}{rgb}{0.4940, 0.1840, 0.5560}
\definecolor{matlabgreen}{rgb}{0.4660, 0.6740, 0.1880}
\definecolor{matlablightblue}{rgb}{0.3010, 0.7450, 0.9330}
\definecolor{matlabred}{rgb}{0.6350, 0.0780, 0.1840}

\begin{tikzpicture}

\begin{axis}[%
width=0.951\fwidth,
height=0.98\fheight,
at={(0\fwidth,0\fheight)},
scale only axis,
xlabel style={font=\color{white!15!black}},
xlabel={$X = L_p$ [dB]},
xmin=0,
xmax=80,
ymin=0,
ymax=1,
ylabel style={font=\color{white!15!black}},
ylabel={$F(X)$},
axis background/.style={fill=white},
axis x line*=bottom,
axis y line*=left,
xmajorgrids,
ymajorgrids,
legend style={
    legend columns=3,
    at={(0.99,0.1)},
    anchor=south east,
    /tikz/every even column/.append style={column sep=1.0cm}
        },
]

\addplot[no markers,matlabblue, line width = 1, dotted] table [x=x, y=y, col sep=comma] {figs/data/Lp_ecdf/elmax47_case_A_Lp_ecdf_rf_130GHz.csv};

\addplot[no markers,matlaborange, line width = 1, dotted] table [x=x, y=y, col sep=comma] {figs/data/Lp_ecdf/elmax47_case_B_Lp_ecdf_rf_130GHz.csv};

\addplot[no markers,matlabyellow, line width = 1, dotted] table [x=x, y=y, col sep=comma] {figs/data/Lp_ecdf/elmax47_case_C_Lp_ecdf_rf_130GHz.csv};

\addplot[no markers,matlabblue, line width = 0.8, dashed] table [x=x, y=y, col sep=comma] {figs/data/Lp_ecdf/elmax47_case_A_Lp_ecdf_rf_220GHz.csv};

\addplot[no markers,matlaborange, line width = 0.8,dashed] table [x=x, y=y, col sep=comma] {figs/data/Lp_ecdf/elmax47_case_B_Lp_ecdf_rf_220GHz.csv};

\addplot[no markers,matlabyellow, line width = 0.8,dashed] table [x=x, y=y, col sep=comma] {figs/data/Lp_ecdf/elmax47_case_C_Lp_ecdf_rf_220GHz.csv};

\addplot[no markers,matlabblue, line width = 0.8] table [x=x, y=y, col sep=comma] {figs/data/Lp_ecdf/elmax47_case_A_Lp_ecdf_rf_660GHz.csv};

\addplot[no markers,matlaborange, line width = 0.8] table [x=x, y=y, col sep=comma] {figs/data/Lp_ecdf/elmax47_case_B_Lp_ecdf_rf_660GHz.csv};

\addplot[no markers,matlabyellow, line width = 0.8] table [x=x, y=y, col sep=comma] {figs/data/Lp_ecdf/elmax47_case_C_Lp_ecdf_rf_660GHz.csv};

\end{axis}
\end{tikzpicture}%
	}

	\hspace{-0.1\columnwidth}\subfloat[Max elevation $\ang{70}$.]{
		\centering
		\definecolor{matlabblue}{rgb}{0.0000, 0.4470, 0.7410}
\definecolor{matlaborange}{rgb}{0.8500, 0.3250, 0.0980}
\definecolor{matlabyellow}{rgb}{0.9290, 0.6940, 0.1250}
\definecolor{matlabpurple}{rgb}{0.4940, 0.1840, 0.5560}
\definecolor{matlabgreen}{rgb}{0.4660, 0.6740, 0.1880}
\definecolor{matlablightblue}{rgb}{0.3010, 0.7450, 0.9330}
\definecolor{matlabred}{rgb}{0.6350, 0.0780, 0.1840}

\begin{tikzpicture}

\begin{axis}[%
width=0.951\fwidth,
height=0.98\fheight,
at={(0\fwidth,0\fheight)},
scale only axis,
xlabel style={font=\color{white!15!black}},
xlabel={$X = L_p$ [dB]},
xmin=0,
xmax=80,
ymin=0,
ymax=1,
ylabel style={font=\color{white!15!black}},
ylabel={$F(X)$},
axis background/.style={fill=white},
axis x line*=bottom,
axis y line*=left,
xmajorgrids,
ymajorgrids,
legend style={
    legend columns=3,
    at={(0.99,0.1)},
    anchor=south east,
    /tikz/every even column/.append style={column sep=1.0cm}
        },
]

\addplot[no markers,matlabblue, line width = 1, dotted] table [x=x, y=y, col sep=comma] {figs/data/Lp_ecdf/elmax70_case_A_Lp_ecdf_rf_130GHz.csv};

\addplot[no markers,matlaborange, line width = 1, dotted] table [x=x, y=y, col sep=comma] {figs/data/Lp_ecdf/elmax70_case_B_Lp_ecdf_rf_130GHz.csv};

\addplot[no markers,matlabyellow, line width = 1, dotted] table [x=x, y=y, col sep=comma] {figs/data/Lp_ecdf/elmax70_case_C_Lp_ecdf_rf_130GHz.csv};

\addplot[no markers,matlabblue, line width = 0.8, dashed] table [x=x, y=y, col sep=comma] {figs/data/Lp_ecdf/elmax70_case_A_Lp_ecdf_rf_220GHz.csv};

\addplot[no markers,matlaborange, line width = 0.8,dashed] table [x=x, y=y, col sep=comma] {figs/data/Lp_ecdf/elmax70_case_B_Lp_ecdf_rf_220GHz.csv};

\addplot[no markers,matlabyellow, line width = 0.8,dashed] table [x=x, y=y, col sep=comma] {figs/data/Lp_ecdf/elmax70_case_C_Lp_ecdf_rf_220GHz.csv};

\addplot[no markers,matlabblue, line width = 0.8] table [x=x, y=y, col sep=comma] {figs/data/Lp_ecdf/elmax70_case_A_Lp_ecdf_rf_660GHz.csv};

\addplot[no markers,matlaborange, line width = 0.8] table [x=x, y=y, col sep=comma] {figs/data/Lp_ecdf/elmax70_case_B_Lp_ecdf_rf_660GHz.csv};

\addplot[no markers,matlabyellow, line width = 0.8] table [x=x, y=y, col sep=comma] {figs/data/Lp_ecdf/elmax70_case_C_Lp_ecdf_rf_660GHz.csv};

\end{axis}
\end{tikzpicture}%
	}

	\hspace{-0.1\columnwidth}\subfloat[Max elevation $\ang{83}$.]{
		\centering
		\definecolor{matlabblue}{rgb}{0.0000, 0.4470, 0.7410}
\definecolor{matlaborange}{rgb}{0.8500, 0.3250, 0.0980}
\definecolor{matlabyellow}{rgb}{0.9290, 0.6940, 0.1250}
\definecolor{matlabpurple}{rgb}{0.4940, 0.1840, 0.5560}
\definecolor{matlabgreen}{rgb}{0.4660, 0.6740, 0.1880}
\definecolor{matlablightblue}{rgb}{0.3010, 0.7450, 0.9330}
\definecolor{matlabred}{rgb}{0.6350, 0.0780, 0.1840}

\begin{tikzpicture}

\begin{axis}[%
width=0.951\fwidth,
height=0.98\fheight,
at={(0\fwidth,0\fheight)},
scale only axis,
xlabel style={font=\color{white!15!black}},
xlabel={$X = L_p$ [dB]},
xmin=0,
xmax=80,
ymin=0,
ymax=1,
ylabel style={font=\color{white!15!black}},
ylabel={$F(X)$},
axis background/.style={fill=white},
axis x line*=bottom,
axis y line*=left,
xmajorgrids,
ymajorgrids,
legend style={
    legend columns=3,
    at={(0.99,0.1)},
    anchor=south east,
    /tikz/every even column/.append style={column sep=1.0cm}
        },
]

\addplot[no markers,matlabblue, line width = 1, dotted] table [x=x, y=y, col sep=comma] {figs/data/Lp_ecdf/elmax83_case_A_Lp_ecdf_rf_130GHz.csv};

\addplot[no markers,matlaborange, line width = 1, dotted] table [x=x, y=y, col sep=comma] {figs/data/Lp_ecdf/elmax83_case_B_Lp_ecdf_rf_130GHz.csv};

\addplot[no markers,matlabyellow, line width = 1, dotted] table [x=x, y=y, col sep=comma] {figs/data/Lp_ecdf/elmax83_case_C_Lp_ecdf_rf_130GHz.csv};

\addplot[no markers,matlabblue, line width = 0.8, dashed] table [x=x, y=y, col sep=comma] {figs/data/Lp_ecdf/elmax83_case_A_Lp_ecdf_rf_220GHz.csv};

\addplot[no markers,matlaborange, line width = 0.8,dashed] table [x=x, y=y, col sep=comma] {figs/data/Lp_ecdf/elmax83_case_B_Lp_ecdf_rf_220GHz.csv};

\addplot[no markers,matlabyellow, line width = 0.8,dashed] table [x=x, y=y, col sep=comma] {figs/data/Lp_ecdf/elmax83_case_C_Lp_ecdf_rf_220GHz.csv};

\addplot[no markers,matlabblue, line width = 0.8] table [x=x, y=y, col sep=comma] {figs/data/Lp_ecdf/elmax83_case_A_Lp_ecdf_rf_660GHz.csv};

\addplot[no markers,matlaborange, line width = 0.8] table [x=x, y=y, col sep=comma] {figs/data/Lp_ecdf/elmax83_case_B_Lp_ecdf_rf_660GHz.csv};

\addplot[no markers,matlabyellow, line width = 0.8] table [x=x, y=y, col sep=comma] {figs/data/Lp_ecdf/elmax83_case_C_Lp_ecdf_rf_660GHz.csv};

\end{axis}
\end{tikzpicture}%
	}
	\caption{\gls{ecdf} of the pointing loss $L_p$ for the three considered trajectories with maximum altitude $\ang{47}$, $\ang{70}$, and $\ang{83}$, for the three considered velocity profiles A, B, and C (color-coded), and for the three different values of gain over Half Power Beamwidth (HPBW) $G_t / \text{HPBW}$ (coded in the line patterns).}
	\label{fig:lp_ecdf}
\end{figure}
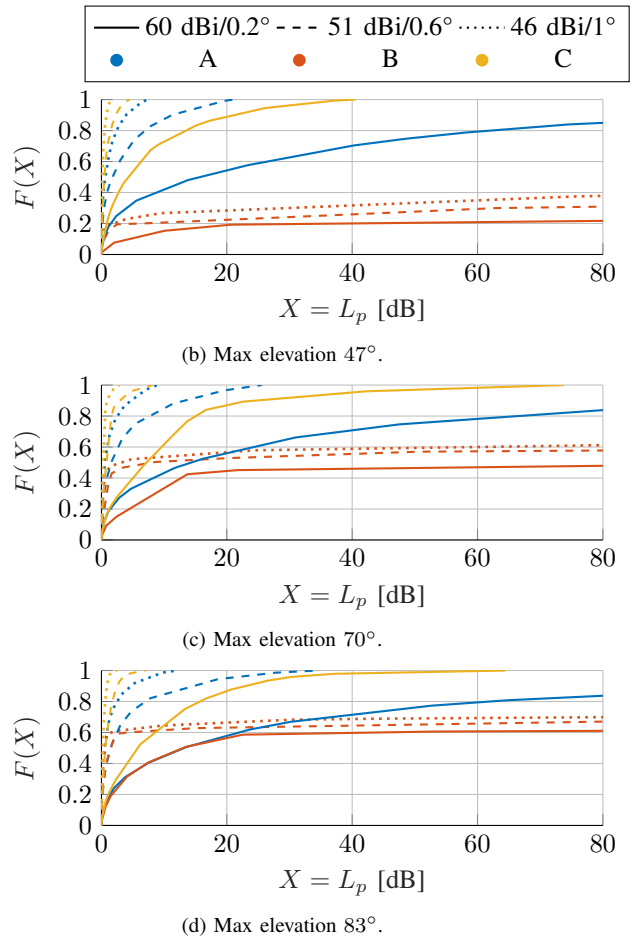

In \cref{fig:lp_ecdf} we report the \gls{ecdf} of the pointing error loss in the considered scenarios.
The pointing loss mirrors the behavior of the pointing error reported in~\cref{fig:pe_ecdf}: for all the considered antennas, the optimized profile consistently results in a lower loss than A and B.
Furthermore, as expected, narrower beams are more susceptible to pointing error.
Nevertheless, profile C maintains the loss below 5~dB for all antenna configurations.

Besides the pointing loss amplitude, a critical aspect is its oscillatory nature.
As analyzed in~\cref{sec:pe_analysis}, sending control commands in regular time intervals results in discrete movements of the antenna, resulting in a variation in pointing loss at each time step.
The frequency of the pointing loss oscillation thus corresponds roughly to the command interval time, and the amplitude depends on the pointing error variation during the same interval.
To measure this effect, we introduce the pointing loss \gls{roc} $R(s)$, computed over a sliding window of duration $W$ seconds and a sliding step of $S$ seconds:
\begin{align}
	R(s) = \frac{\max(L_p(W_s))-\min(L_p(W_s))}{W}, & s\in\mathbb{N},
\end{align}
where $W_s$ represents the window at time $s$ and $L_p(W_s)$ the pointing loss observed during $W_s$.

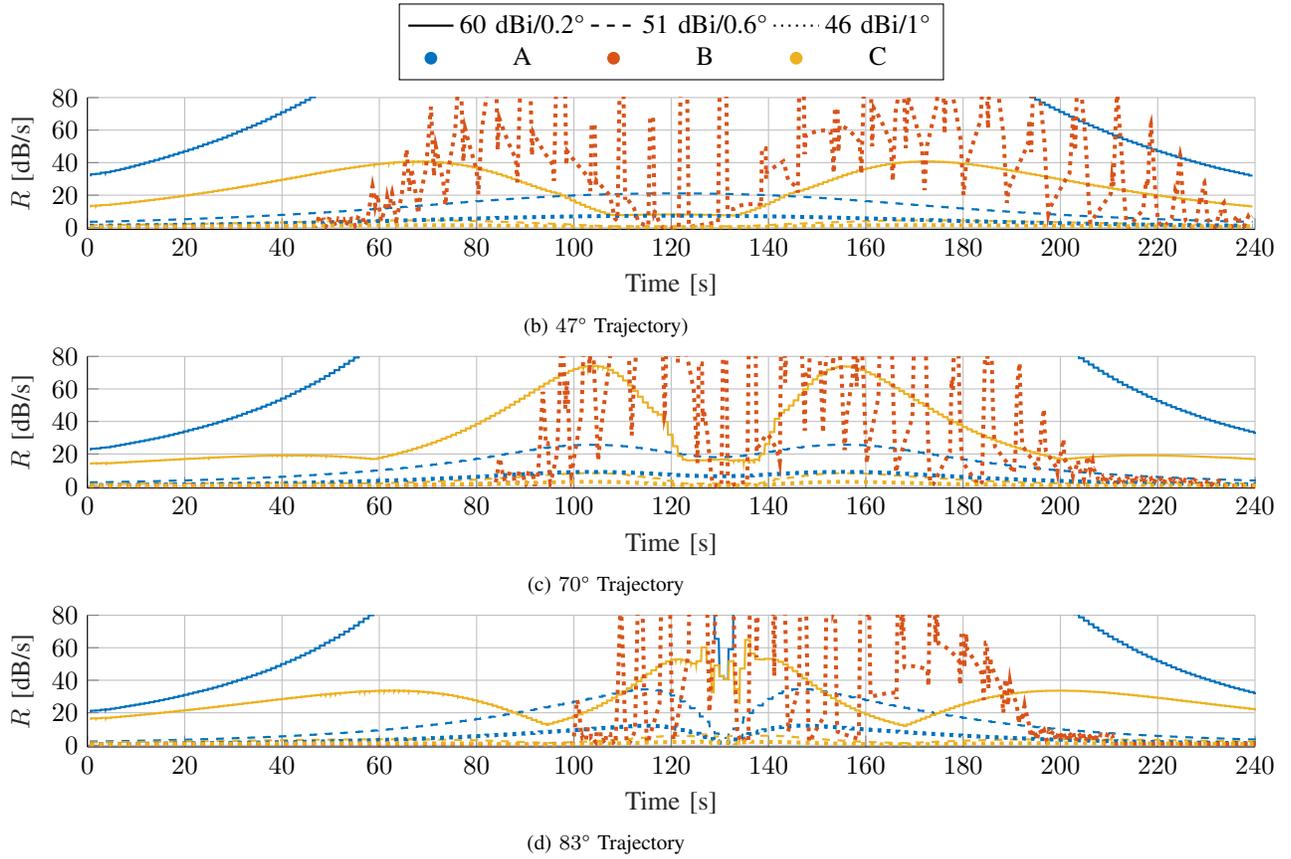
\begin{figure*}[t]
	\centering
	\subfloat{
		\centering
		\definecolor{matlabblue}{rgb}{0.0000, 0.4470, 0.7410}
\definecolor{matlaborange}{rgb}{0.8500, 0.3250, 0.0980}
\definecolor{matlabyellow}{rgb}{0.9290, 0.6940, 0.1250}
\definecolor{matlabpurple}{rgb}{0.4940, 0.1840, 0.5560}
\definecolor{matlabgreen}{rgb}{0.4660, 0.6740, 0.1880}
\definecolor{matlablightblue}{rgb}{0.3010, 0.7450, 0.9330}
\definecolor{matlabred}{rgb}{0.6350, 0.0780, 0.1840}

\begin{tikzpicture}

\begin{axis}[%
width=0.9125\fwidth,
height=0.9125\fheight,
at={(0\fwidth,0\fheight)},
scale only axis,
hide axis,
legend style={
    legend columns=3,
    /tikz/every even column/.append style={column sep=0mm}
        },
    xmin=10,
    xmax=50,
    ymin=0,
    ymax=0.4,
]

\addlegendimage{black, solid, line width = 0.8}
\addlegendentry{60 dBi/0.2$\degree$}
\addlegendimage{black, dashed, line width = 0.8}
\addlegendentry{51 dBi/0.6$\degree$}
\addlegendimage{black, dotted, line width = 0.8}
\addlegendentry{46 dBi/1$\degree$}
\addlegendimage{matlabblue, solid, line width = 0.8,  only marks}
\addlegendentry{A}
\addlegendimage{matlaborange, solid, line width = 0.8, only marks}
\addlegendentry{B}
\addlegendimage{matlabyellow, solid, line width = 0.8, only marks}
\addlegendentry{C}
\end{axis}

\end{tikzpicture}%
	}

	\setlength{\fheight}{0.2\columnwidth}
	\hspace{-0.1\columnwidth}\subfloat[$47\degree$ Trajectory)\label{fig:roc_47}]{
		\setlength{\fwidth}{0.9\textwidth}
		\centering
		\definecolor{matlabblue}{rgb}{0.0000, 0.4470, 0.7410}
\definecolor{matlaborange}{rgb}{0.8500, 0.3250, 0.0980}
\definecolor{matlabyellow}{rgb}{0.9290, 0.6940, 0.1250}
\definecolor{matlabpurple}{rgb}{0.4940, 0.1840, 0.5560}
\definecolor{matlabgreen}{rgb}{0.4660, 0.6740, 0.1880}
\definecolor{matlablightblue}{rgb}{0.3010, 0.7450, 0.9330}
\definecolor{matlabred}{rgb}{0.6350, 0.0780, 0.1840}

\begin{tikzpicture}

\begin{axis}[%
width=0.951\fwidth,
height=\fheight,
at={(0\fwidth,0\fheight)},
scale only axis,
xlabel style={font=\color{white!15!black}},
xlabel={Time [s]},
xmin=0,
xmax=240,
ymin = -1,
ymax = 80,
ylabel style={font=\color{white!15!black}},
ylabel={$R$ [dB/s]},
axis background/.style={fill=white, draw=none},
axis x line*=bottom,
axis y line*=left,
xmajorgrids,
ymajorgrids,
legend style={
    legend columns=3,
    at={(0.99,0.1)},
    anchor=south east,
    /tikz/every even column/.append style={column sep=1.0cm}
        },
]

\addplot[no markers,matlabblue, solid, line width = 0.8] table [x=x, y=y, col sep=comma] {figs/data/t_vs_roc/elmax47_case_A_t_vs_roc_rf_660GHz.csv};

\addplot[no markers,matlabyellow, solid, line width = 0.8] table [x=x, y=y, col sep=comma] {figs/data/t_vs_roc/elmax47_case_C_t_vs_roc_rf_660GHz.csv};

\addplot[no markers,matlabblue, dashed, line width = 0.8] table [x=x, y=y, col sep=comma] {figs/data/t_vs_roc/elmax47_case_A_t_vs_roc_rf_220GHz.csv};

\addplot[no markers,matlabyellow, dashed, line width = 0.8] table [x=x, y=y, col sep=comma] {figs/data/t_vs_roc/elmax47_case_C_t_vs_roc_rf_220GHz.csv};

\addplot[no markers,matlabblue, dotted, line width = 1.5] table [x=x, y=y, col sep=comma] {figs/data/t_vs_roc/elmax47_case_A_t_vs_roc_rf_130GHz.csv};

\addplot[no markers,matlaborange, dotted, line width = 1.5] table [x=x, y=y, col sep=comma] {figs/data/t_vs_roc/elmax47_case_B_t_vs_roc_rf_130GHz.csv};

\addplot[no markers,matlabyellow, dotted, line width = 1.5] table [x=x, y=y, col sep=comma] {figs/data/t_vs_roc/elmax47_case_C_t_vs_roc_rf_130GHz.csv};

\end{axis}
\end{tikzpicture}%
	}

	\hspace{-0.1\columnwidth}\subfloat[$70\degree$ Trajectory\label{fig:roc_70}]{
		\setlength{\fwidth}{0.9\textwidth}
		\centering
		\definecolor{matlabblue}{rgb}{0.0000, 0.4470, 0.7410}
\definecolor{matlaborange}{rgb}{0.8500, 0.3250, 0.0980}
\definecolor{matlabyellow}{rgb}{0.9290, 0.6940, 0.1250}
\definecolor{matlabpurple}{rgb}{0.4940, 0.1840, 0.5560}
\definecolor{matlabgreen}{rgb}{0.4660, 0.6740, 0.1880}
\definecolor{matlablightblue}{rgb}{0.3010, 0.7450, 0.9330}
\definecolor{matlabred}{rgb}{0.6350, 0.0780, 0.1840}

\begin{tikzpicture}

\begin{axis}[%
width=0.951\fwidth,
height=\fheight,
at={(0\fwidth,0\fheight)},
scale only axis,
xlabel style={font=\color{white!15!black}},
xlabel={Time [s]},
xmin=0,
xmax=240,
ymin = -1,
ymax = 80,
ylabel style={font=\color{white!15!black}},
ylabel={$R$ [dB/s]},
axis background/.style={fill=white},
axis x line*=bottom,
axis y line*=left,
xmajorgrids,
ymajorgrids,
legend style={
    legend columns=3,
    at={(0.99,0.1)},
    anchor=south east,
    /tikz/every even column/.append style={column sep=1.0cm}
        },
]

\addplot[no markers,matlabblue, dashed, line width = 0.8] table [x=x, y=y, col sep=comma] {figs/data/t_vs_roc/elmax70_case_A_t_vs_roc_rf_220GHz.csv};

\addplot[no markers,matlabyellow, dashed, line width = 0.8] table [x=x, y=y, col sep=comma] {figs/data/t_vs_roc/elmax70_case_C_t_vs_roc_rf_220GHz.csv};

\addplot[no markers,matlabblue, solid, line width = 0.8] table [x=x, y=y, col sep=comma] {figs/data/t_vs_roc/elmax70_case_A_t_vs_roc_rf_660GHz.csv};

\addplot[no markers,matlabyellow, solid, line width = 0.8] table [x=x, y=y, col sep=comma] {figs/data/t_vs_roc/elmax70_case_C_t_vs_roc_rf_660GHz.csv};

\addplot[no markers,matlabblue, dotted, line width = 1.5] table [x=x, y=y, col sep=comma] {figs/data/t_vs_roc/elmax70_case_A_t_vs_roc_rf_130GHz.csv};

\addplot[no markers,matlaborange, dotted, line width = 1.5] table [x=x, y=y, col sep=comma] {figs/data/t_vs_roc/elmax70_case_B_t_vs_roc_rf_130GHz.csv};

\addplot[no markers,matlabyellow, dotted, line width = 1.5] table [x=x, y=y, col sep=comma] {figs/data/t_vs_roc/elmax70_case_C_t_vs_roc_rf_130GHz.csv};

\end{axis}
\end{tikzpicture}%
	}

	\hspace{-0.1\columnwidth}\subfloat[$83\degree$ Trajectory\label{fig:roc_83}]{
		\setlength{\fwidth}{0.9\textwidth}
		\centering
		\definecolor{matlabblue}{rgb}{0.0000, 0.4470, 0.7410}
\definecolor{matlaborange}{rgb}{0.8500, 0.3250, 0.0980}
\definecolor{matlabyellow}{rgb}{0.9290, 0.6940, 0.1250}
\definecolor{matlabpurple}{rgb}{0.4940, 0.1840, 0.5560}
\definecolor{matlabgreen}{rgb}{0.4660, 0.6740, 0.1880}
\definecolor{matlablightblue}{rgb}{0.3010, 0.7450, 0.9330}
\definecolor{matlabred}{rgb}{0.6350, 0.0780, 0.1840}

\begin{tikzpicture}

\begin{axis}[%
width=0.951\fwidth,
height=\fheight,
at={(0\fwidth,0\fheight)},
scale only axis,
xlabel style={font=\color{white!15!black}},
xlabel={Time [s]},
xmin=0,
xmax=240,
ymin = -1,
ymax = 80,
ylabel style={font=\color{white!15!black}},
ylabel={$R$ [dB/s]},
axis background/.style={fill=white},
axis x line*=bottom,
axis y line*=left,
xmajorgrids,
ymajorgrids,
legend style={
    legend columns=3,
    at={(0.99,0.1)},
    anchor=south east,
    /tikz/every even column/.append style={column sep=1.0cm}
        },
]

\addplot[no markers,matlabblue, dashed, line width = 0.8] table [x=x, y=y, col sep=comma] {figs/data/t_vs_roc/elmax83_case_A_t_vs_roc_rf_220GHz.csv};

\addplot[no markers,matlabyellow, dashed, line width = 0.8] table [x=x, y=y, col sep=comma] {figs/data/t_vs_roc/elmax83_case_C_t_vs_roc_rf_220GHz.csv};

\addplot[no markers,matlabblue, solid, line width = 0.8] table [x=x, y=y, col sep=comma] {figs/data/t_vs_roc/elmax83_case_A_t_vs_roc_rf_660GHz.csv};

\addplot[no markers,matlabyellow, solid, line width = 0.8] table [x=x, y=y, col sep=comma] {figs/data/t_vs_roc/elmax83_case_C_t_vs_roc_rf_660GHz.csv};

\addplot[no markers,matlabblue, dotted, line width = 1.5] table [x=x, y=y, col sep=comma] {figs/data/t_vs_roc/elmax83_case_A_t_vs_roc_rf_130GHz.csv};

\addplot[no markers,matlaborange, dotted, line width = 1.5] table [x=x, y=y, col sep=comma] {figs/data/t_vs_roc/elmax83_case_B_t_vs_roc_rf_130GHz.csv};

\addplot[no markers,matlabyellow, dotted, line width = 1.5] table [x=x, y=y, col sep=comma] {figs/data/t_vs_roc/elmax83_case_C_t_vs_roc_rf_130GHz.csv};

\end{axis}
\end{tikzpicture}%
	}
	\caption{\gls{roc} of the pointing loss in the three considered transits, with different velocity profiles and antenna gains. The narrower the beam, the more pronounced the pointing loss. The \gls{roc} is computed with a sliding window of 1~s, with a 5~ms step. We report the \gls{roc} obtained with velocity profile B only for the broadest beam, for clarity. As expected, the performance with narrower beams further decreases.}
	\vspace{-0.2in}
	\label{fig:roc}
\end{figure*}

In \cref{fig:roc}, we report the \gls{roc} for the considered scenarios.
Velocity profile B exhibits a clear oscillatory behavior even in terms of \gls{roc}, which peaks well above 80 dB/s for all the considered trajectories.
On the contrary, other velocity profiles show limited perturbations, except for the  $0.2\degree$ beam, for which  profile A observes substantial pointing error losses.
Similarly, amplitude of the \gls{roc} of profile C reports a significant and frequent fading for the most narrow beam, but falls within an acceptable 5~dB/s variation rate when using a larger beam.

\section{Conclusion}
In this work, we study the open-loop tracking of a \gls{leo} satellite for \gls{sthz}/\gls{thz} uplink, focusing on the mathematical model that captures the physical and control-related dynamics of a real tracking system, on quantifying the pointing error over realistic satellite passes, and incorporating these effects into the link budget, thus connecting hardware-level limitations directly to link-level performance metrics.

This work analyzes the sensitivity of system performance due to high directional beams, a problem which is amplified by the rapid apparent motion of the satellite relative to the ground terminal, the lack of instantaneous feedback in some operational modes, such as open-loop pointing.
We show an evaluation of the trade-offs between beam directionality and pointing tolerance, supported by performance results for multiple pass geometries and
mount control strategies.
We also provide insights for the design of robust high-frequency \gls{ntn} uplink, where precise beam alignment must be achieved under practical mechanical constraints.

Future work could include optimizing the time step and its duration, based on the hardware characteristics, relaxing the hardware constrains to dynamically tune the velocity and acceleration at each step, accounting for atmospheric effects, and studying the effect of the pointing-induced fading on higher layers and end-to-end performance.

\section*{Acknowledgment}
This work was supported in part by the U.S. National Science Foundation Grants CNS-2332721 and CNS-2346487.

\bibliographystyle{IEEEtran}
\bibliography{references}

\end{document}